\documentclass[aps,nofootinbib,preprintnumbers]{revtex4}

\usepackage{graphicx}
\usepackage{graphics}
\usepackage{amssymb}
\usepackage{amsmath}
\usepackage{bm}   

\newcommand{\nslash}{\kern 0.2 em n\kern -0.50em /}

\begin{document}

\title{Single-spin asymmetries: the Trento conventions}

\author{Alessandro Bacchetta}
\email{alessandro.bacchetta@physik.uni-r.de}
\affiliation{
Institut f{\"u}r Theoretische Physik, Universit{\"a}t Regensburg,
D-93040 Regensburg, Germany
}

\author{Umberto D'Alesio}
\email{umberto.dalesio@ca.infn.it}
\affiliation{
INFN, Sezione di Cagliari and Dipartimento di Fisica, 
Universit{\`a} di Cagliari,
I-09042 Monserrato, Italy
}

\author{Markus Diehl}
\email{markus.diehl@desy.de}
\affiliation{
Deutsches Elektronen-Synchroton DESY, D-22603 Hamburg, Germany
}

\author{C. Andy Miller}
\email{miller@triumf.ca}
\affiliation{
TRIUMF, Vancouver, British Columbia V6T 2A3, Canada
}

\begin{abstract}
During the workshop ``Transversity:~New Developments in Nucleon Spin
Structure'' (ECT$^\ast$, Trento, Italy, 14--18 June 2004), a series of
recommendations was put forward by the participants concerning
definitions and notations for describing effects of intrinsic
transverse momentum of partons in semi-inclusive deep inelastic
scattering.
\end{abstract}

\preprint{DESY 04-193}

\maketitle

\section{Definition of transverse-momentum dependent functions}
\label{s:functions}

A standard set of definitions and notations for transverse-momentum
dependent distribution and fragmentation functions is given in
Refs.~\cite{Mulders:1996dh,Boer:1998nt,Boer:1999uu}.  We note that the
definition of the antisymmetric tensor in those articles and in the
present note is such that
\begin{equation}
\epsilon^{0123}=+1.
\end{equation}

Transverse-momentum dependent parton distributions of leading twist
can be interpreted as number densities (see
e.g.\ Refs.~\cite{Anselmino:1995tv,Anselmino:1999pw,Barone:2001sp}).  To
connect with this interpretation, we take the example of the
distribution of unpolarized quarks in a polarized proton, which is
given by\footnote{The following expression is obtained from the quark
correlation function in Eq.~(2) of Ref.~\protect\cite{Boer:1998nt} by
identifying $n=n_-$, 
multiplying with $\nslash /2$ and taking the
trace.}
\begin{equation}
\begin{split}
f_{q/p^\uparrow}(x, {k}_T) &= f_1^q (x, k_T^2) -
f_{1T}^{\perp q}(x, k_T^2)\,\frac{\epsilon^{\mu \nu \rho \sigma}
  P_{\mu} k_{\nu} S_{\rho} n_{\sigma}
   }{M \; (P \cdot n)} \\
 &= f_1^q (x, k_T^2) -
f_{1T}^{\perp q}(x, k_T^2)\, \frac{(\hat{\bm P} \times {\bm k}_T) 
  \cdot {\bm S}}{M} ,
\end{split}
  \label{sivers-def}
\end{equation}
where $f_1^q$ is the unpolarized quark density and $f_{1T}^{\perp q}$
describes the Sivers effect~\cite{Sivers:1990cc}.  Here $P$ is the
momentum of the proton, $S$ is its covariant spin vector normalized to
$S^2 = -1$, and $M$ is the proton mass.  The covariant definition of
parton distributions requires an auxiliary lightlike vector $n$, 
which plays the role of a
preferred direction in a given physical process.\footnote{ This direction
can for instance be taken along the virtual photon momentum in deep
inelastic scattering, or along the momentum of the second incoming hadron
in Drell-Yan lepton pair production.  Other choices of $n$ are possible,
provided that the corresponding changes in the result are sufficiently
suppressed by
inverse powers of the large momentum scale.}
Furthermore, $k$ is the
momentum of the quark, $k_T$ its component perpendicular to $P$ and
$n$, and $x = (k \cdot n)/(P \cdot n)$ its light-cone momentum
fraction.  The second expression in (\ref{sivers-def}) holds in any
frame where ${\bm n}$ and the direction $\hat{\bm P}$ of the proton
momentum point in opposite directions.\footnote{We use the four-vector
$k_T$ and its square as arguments in the distribution functions to
emphasize that they are Lorentz invariant.  One may instead use ${\bm
k}_T$ if it is clear from the context to which frame the vectors
refer.}  Therefore $f_{1T}^{\perp q} > 0$ corresponds to a preference of
the quark to move to the left if the proton is moving towards the
observer and the proton spin is pointing upwards.  In the convention
of Ref.~\cite{Anselmino:2002pd} the Sivers effect is described by
\begin{equation}
f_{q/p^\uparrow}(x, {k}_T) - f_{q/p^\uparrow}(x, -{k}_T)
= \Delta^N f_{q/p^\uparrow}(x, k_T^2) \, \frac{(\hat{\bm P}
  \times {\bm k}_T) \cdot {\bm S}}{|{\bm k}_T|}
\end{equation}
so that
\begin{equation}
\Delta^N f_{q/p^\uparrow}(x, k_T^2) = 
{}- \frac{2 |{\bm k}_T|}{M}\, f_{1T}^{\perp q}(x, k_T^2) .
\label{sivers-rel}
\end{equation}
Either $f_{1T}^{\perp q}$ or $\Delta^N f_{q/p^\uparrow}$ may be
referred to as the ``Sivers function''.  It is strongly encouraged that
authors use one or the other of these notations, or provide the
relation of the functions they might use to the ones discussed here.

Let us give the corresponding relation for the Boer-Mulders function,
introduced 
in Ref.~\cite{Boer:1998nt}. The distribution of transversely
polarized quarks in an unpolarized proton is\footnote{The following expression is obtained by identifying
$n=n_-$, setting $S_T$ and $\lambda$ to zero, 
multiplying Eq.~(2) in Ref.~\protect\cite{Boer:1998nt} with
$\gamma^\mu n_\mu /2 + i \sigma_{\mu\nu} \gamma_5\, n^\mu
S_{q}^{\nu}/2$, taking the trace and dividing by~2.  See Eq.~(11) and (12)
of~\cite{Boglione:1999pz} for this connection to the number
density interpretation.}
\begin{equation}
\begin{split}
f_{q^\uparrow/p}(x, {k}_T) &= \frac{1}{2}\left(f_1^q (x, k_T^2) -
h_{1}^{\perp q}(x, k_T^2)\,\frac{\epsilon^{\mu \nu \rho \sigma}
  P_{\mu} k_{\nu} S_{q \rho} n_{\sigma}
   }{M \; (P \cdot n)}\right) \\
 &= \frac{1}{2}\left(f_1^q (x, k_T^2) -
h_{1}^{\perp q}(x, k_T^2)\, \frac{(\hat{\bm P} \times {\bm k}_T) 
  \cdot {\bm S}_q}{M}\right) ,
\end{split}
  \label{boermulders-def}
\end{equation}
where $S_q$ is the covariant spin vector of the quark. 
Introducing
\begin{equation}
f_{q^\uparrow/p}(x, {k}_T) - f_{q^\uparrow/p}(x, -{k}_T)
= \Delta^N f_{q^\uparrow/p}(x, k_T^2) \, \frac{(\hat{\bm P}
  \times {\bm k}_T) \cdot {\bm S}_q}{|{\bm k}_T|}
\end{equation}
we get the relation 
\begin{equation}
\Delta^N f_{q^\uparrow/p}(x, k_T^2) = 
{}- \frac{|{\bm k}_T|}{M}\, h_{1}^{\perp q}(x, k_T^2) .
\label{boermulders-rel}
\end{equation}

Likewise there are two common notations for the Collins fragmentation
function~\cite{Collins:1993kk}.  With the conventions of
Refs.~\cite{Mulders:1996dh,Boer:1998nt,Boer:1999uu} the number density
of an unpolarized hadron $h$ in a transversely polarized quark is
\footnote{The following expression is obtained by identifying
$n'=n_+$, setting $S_{hT}$ and $\lambda_h$ to zero, 
multiplying Eq.~(5) in Ref.~\protect\cite{Boer:1998nt} with
$\gamma^\mu n'_\mu /2 + i \sigma_{\mu\nu} \gamma_5\, n'^\mu
S_{q}^{\nu}/2$ and taking the trace.  See Eqs.~(40) and (41)
of~\cite{Boglione:1999pz}.}
\begin{equation}
\begin{split}
D_{h/q^\uparrow}(z,{P}_{hT}) &= D_1^q(z,P_{hT}^2)
 - H_1^{\perp q}(z, P_{hT}^2) \, \frac{\epsilon^{\mu\nu\rho\sigma}
   P_{h \mu} k_\nu S_{q \rho} n'{}_{\!\!\sigma}}{M_h (P_h \cdot n')} \\
 &= D_1^q(z, P_{hT}^2) + H_1^{\perp q}(z, P_{hT}^2) \,
    \frac{(\hat{\bm k} \times {\bm P}_{hT}) \cdot {\bm S}_q}{z M_h},
\end{split}
  \label{collins-def}
\end{equation}
where the measure of the density is $dz\, d^2 P_{hT}$.  Here $D_1^q$ is
the unpolarized fragmentation function, $P_h$ is the hadron momentum,
$M_h$ its mass, $k$ is the momentum of the quark, $S_q$ its
covariant spin vector,
and $n'$ an auxiliary lightlike vector. Furthermore, 
$z = (P_h \cdot n')/(k \cdot n')$ is the light-cone momentum
fraction of the hadron with respect to the fragmenting quark, and
$P_{hT}$ the component of $P_h$ transverse to $k$ and
$n'$.  One can trade $P_{hT}$ for $k_T = -
P_{hT} /z$, the component of $k$ transverse to $P_h$ and $n'$.  The second line of
(\ref{collins-def}) holds in frames where ${\bm n'}$ and the direction
$\hat{\bm k}$ of the quark momentum point in opposite directions.
Therefore, $H_1^{\perp q} > 0$ corresponds to a preference of the
hadron to move to the left if the quark is moving away from the
observer and the quark spin is pointing upwards.  In the notation
of~\cite{Anselmino:2000mb} the Collins effect is described by
\begin{equation}
D_{h/q^\uparrow}(z,{P}_{hT}) - D_{h/q^\uparrow}(z,-{P}_{hT})
 = \Delta^N D_{h/q^{\uparrow}}(z,P_{hT}^2) \,
   \frac{(\hat{\bm k} \times {\bm P}_{hT}) \cdot 
   {\bm S}_q}{|{\bm P}_{hT}|}
\end{equation}
so that 
\begin{equation}
\Delta^N D_{h/q^{\uparrow}}(z,P_{hT}^2) =
\frac{2 |{\bm P}_{hT}|}{z M_h} \,  H_1^{\perp q}(z, P_{hT}^2) .
\label{collins-rel}
\end{equation} 
Either $H_1^{\perp q}$ or $\Delta^N D_{h/q^\uparrow}$ may be referred to as
``Collins function''.
Our relations (\ref{sivers-rel}), (\ref{boermulders-rel}), (\ref{collins-rel}) agree with (4.8.3a), (4.8.3b), (6.5.11)
in Ref.~\cite{Barone:2001sp}.

We finally discuss the analog of the Sivers function in fragmentation, 
introduced by Mulders and Tangerman in
Ref.~\cite{Mulders:1996dh}. The number density
of a polarized spin-half hadron $h$ in an unpolarized quark is\footnote{The following expression is obtained by identifying
$n'=n_+$, multiplying Eq.~(5) in Ref.~\protect\cite{Boer:1998nt} with
$\nslash'/2$, taking the trace and dividing by~2.}
\begin{equation}
\begin{split}
D_{h^\uparrow/q}(z,{P}_{hT}) &= \frac{1}{2} \left(D_1^q(z,P_{hT}^2)
 - D_{1T}^{\perp q}(z, P_{hT}^2) \, \frac{\epsilon^{\mu\nu\rho\sigma}
   P_{h \mu} k_\nu S_{h \rho} n'{}_{\!\!\sigma}}{M_h (P_h \cdot n')}\right) \\
 &= \frac{1}{2} \left( D_1^q(z, P_{hT}^2) + D_{1T}^{\perp q}(z, P_{hT}^2) \,
    \frac{(\hat{\bm k} \times {\bm P}_{hT}) \cdot {\bm S}_h}{z M_h}\right),
\end{split}
  \label{multan-def}
\end{equation}
where $S_h$ is the covariant spin vector of the hadron.
As indicated in Ref.~\cite{Anselmino:2001js}, we can write
\begin{equation}
D_{h^\uparrow/q}(z,{P}_{hT}) - D_{h^\uparrow/q}(z,-{P}_{hT})
 = \Delta^N D_{h^\uparrow/q}(z,P_{hT}^2) \,
   \frac{(\hat{\bm k} \times {\bm P}_{hT}) \cdot 
   {\bm S}_h}{|{\bm P}_{hT}|},
\end{equation}
which leads 
to\footnote{Note that there is a factor $-2$ 
too much in Eq.~(5) of Ref.~\cite{Anselmino:2001js}. 
This does not affect any results in that work.}
\begin{equation}
\Delta^N D_{h^{\uparrow}/q}(z,P_{hT}^2) =
\frac{|{\bm P}_{hT}|}{z M_h} \,  D_{1T}^{\perp q}(z, P_{hT}^2) .
\end{equation} 

The definition of each parton distribution contains a Wilson line, which
describes interactions with the spectator partons before or after the
hard-scattering process.  The path of this Wilson line in space-time
is selected by the hard process in which the parton distribution
appears.  Each such path corresponds to its own set of distribution
functions, which thus give the number of quarks found in the
presence of the specified spectator interactions.  Different paths can
lead to different distributions, and the path should be
specified in the notation when it is not evident from the
context.\footnote{This has been realized only recently, and the
necessary distinction is not made in
\protect\cite{Sivers:1990cc,Collins:1993kk,Anselmino:1995tv,Mulders:1996dh,Boer:1998nt,Boer:1999uu,Boglione:1999pz,Anselmino:1999pw,Anselmino:2000mb,Barone:2001sp,Anselmino:2001js,Anselmino:2002pd}.}
Using time reversal symmetry one can show~\cite{Collins:2002kn}
\begin{equation}
f_1^{\rm DIS}(x,k_T^2) = f_1^{\rm DY}(x,k_T^2) , \qquad
f_{1T}^{\perp {\rm DIS}}(x,k_T^2) = - f_{1T}^{\perp \rm{DY}}(x,k_T^2) ,
\end{equation}
where the superscripts respectively specify the distributions with
Wilson lines appropriate for semi-inclusive deep inelastic scattering
(SIDIS) and for Drell-Yan lepton pair production.

Wilson lines with a path selected by the process also appear in the
definition of fragmentation functions.  The relation between the
functions relevant for different processes (such as $e^+e^-$
annihilation or SIDIS) is currently under study.

\section{Azimuthal angles in semi-inclusive deep inelastic scattering}

\begin{figure}[ht] 
\includegraphics{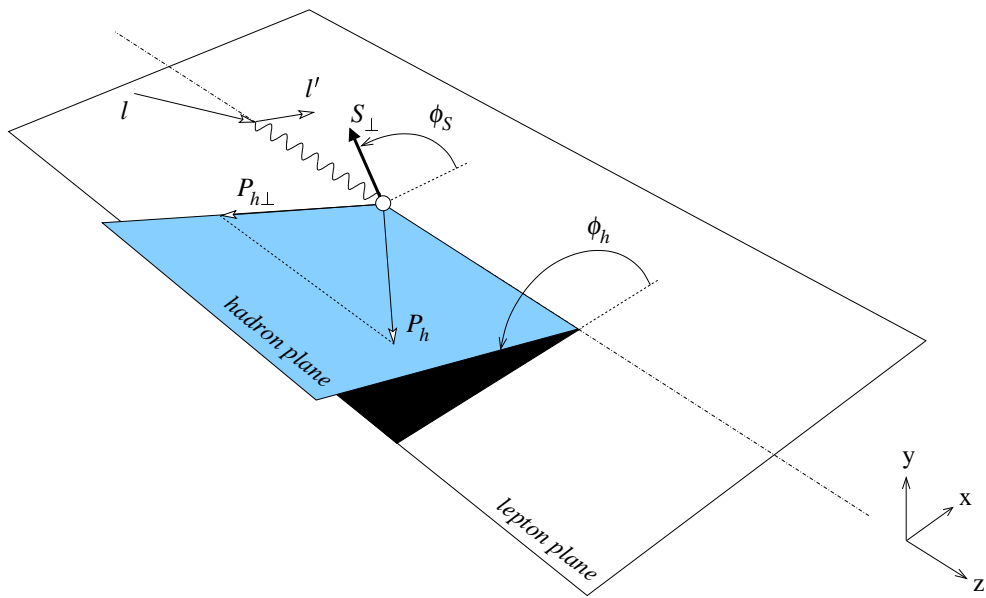}
\caption{Definition of azimuthal angles for the process (\ref{sidis})
  in the target rest frame.  $P_{h\perp}$ and $S_\perp$ are the
  components of $P_h$ and $S$ transverse to the photon
  momentum.}
\label{f:angles}
\end{figure}

A recommendation is made concerning the azimuthal angles
relevant in the semi-inclusive cross section for
\begin{equation}
  \label{sidis}
\ell(l) + p(P) \to \ell(l') + h(P_h) + X ,
\end{equation}
where $\ell$ denotes the beam lepton, $p$ the proton target, and $h$
the produced hadron. As usual we define $q = l - l'$ and
$Q^2 = - q^2$.   The azimuthal angle $\phi_h$ between the lepton
and the hadron planes should be defined as
\begin{align}
  \label{angle-def-1}
\begin{split}
\cos \phi_h &= 
  \frac{(\hat{\bm q}\times{\bm l})}{|\hat{\bm q}\times{\bm l}|}
  \cdot \frac{(\hat{\bm q}\times{\bm P}_h)}{|\hat{\bm q}
     \times{\bm P}_h|},\\
\sin \phi_h &= 
  \frac{({\bm l} \times {\bm P}_h) \cdot \hat{\bm q}}{|\hat{\bm q}
     \times{\bm l}|\,|\hat{\bm q}\times{\bm P}_h|} , 
\end{split}
\end{align}
with $\hat{\bm q}={\bm q}/|{\bm q}|$, where all vectors refer to the
target rest frame (or to any frame reached from the target rest frame
by a boost along $\hat{\bm q}$).  Writing the right-hand sides of
(\ref{angle-def-1}) in a Lorentz invariant form, one has
\begin{align}
  \label{angle-def-2}
\begin{split}
\cos \phi_h &=  {}- \frac{g_{\perp}^{\mu \nu}l_{\mu} P_{h \nu}}{
  |l_\perp|\, |P_{h \perp}|} , \\
\sin \phi_h &= {}- \frac{\epsilon_{\perp}^{\mu \nu}l_{\mu} P_{h \nu}}{
  |l_\perp|\, |P_{h \perp}|}
\end{split}
\end{align}
with $|l_\perp| = \sqrt{-g_{\perp}^{\mu \nu} l_{\mu} l_{\nu}}$ and
$|P_{h \perp}| = \sqrt{-g_{\perp}^{\mu \nu} P_{h \mu} P_{h \nu}}$.
Here we introduced perpendicular projection tensors
\begin{align}
  \label{perp-tensors}
\begin{split}
g_{\perp}^{\mu \nu} &= g^{\mu \nu} - 
\frac{q^{\mu} P^{\nu} + P^{\mu} q^{\nu}}{P \cdot q \;(1+\gamma^2)}
+ \frac{\gamma^2}{1+\gamma^2}
  \left(\frac{q^{\mu} q^{\nu}}{Q^2}-\frac{P^{\mu}P^{\nu}}{M^2} \right), \\ 
\epsilon_{\perp}^{\rho \sigma} &= \epsilon^{\mu \nu \rho \sigma}
  \frac{{P}_{\mu} {q}_{\nu}}{P\cdot q \;\sqrt{1+\gamma^2}}
\end{split}
\end{align}
with $\gamma = 2 x M/Q$, where $x$ is the Bjorken variable and $M$
again the target mass.  Evaluating the right-hand sides
of~(\ref{angle-def-2}) in the target rest frame, one recovers
(\ref{angle-def-1}).  The azimuthal angle $\phi_S$ relevant for
specifying the target polarization is defined in analogy to
(\ref{angle-def-1}) and (\ref{angle-def-2}), with $P_h$ replaced by
the covariant spin vector $S$ of the target.  The definitions of
$\phi_h$ and $\phi_S$ are illustrated in Fig.~\ref{f:angles}.  We
emphasize that (\ref{angle-def-1}), (\ref{angle-def-2}),
(\ref{perp-tensors}) do {\em not} depend on the choice of coordinate
axes.  For definiteness we show in Fig.~\ref{f:angles} one frequently used 
coordinate system. In this system the tensors defined in
Eq.~(\ref{perp-tensors}) have nonzero components $g_{\perp}^{11} =
g_{\perp}^{22} = -1$ and $\epsilon_{\perp}^{12} =
-\epsilon_{\perp}^{21} = -1$. 
Note that two different conventions for drawing angles 
and interpreting their sign in figures
are in general use in the literature:  
\renewcommand{\theenumi}{\Alph{enumi}}
\renewcommand{\labelenumi}{\theenumi.}
\begin{enumerate}
\item{
\label{conv1}
The $z$ axis is specified and angles are drawn as arcs with one arrowhead. 
If an angle is oriented according to the
right-hand rule it is {\em positive}, otherwise it is {\em negative}.
Fig.~\ref{f:angles} illustrates the application of this convention.
}
\item{\label{conv2}
Illustrated angles are always assumed to be
{\em positive}. Only the location of the arc
 affects the definition of the angle. 
No orientation should be assigned 
to the arc, and any $z$ axis that may be present does
not affect the angle definition.
}
\end{enumerate}
It is strongly recommended that authors avoid placing
single arrowheads on arcs when using convention \ref{conv2}.
When using convention \ref{conv1}, an explicit remark in the caption
  may be useful when the figure illustrates a situation in which
  an angle has a negative value.

Theorists often prefer a coordinate system with the same $x$ 
axis but with $y$ and $z$ axes opposite to those shown in
Fig.~\ref{f:angles}, so that in the $\gamma^* p$ center of mass the 
target moves in the positive $z$ direction 
(cf.\ Sect.~\protect\ref{s:functions}). When working in that 
coordinate system in the context of graphical convention \ref{conv1}
one can conform with the 
definition of angles recommended here
by using the opposite orientation for both $\phi_h$ and $\phi_S$.

We note that the angles $\phi_h$ and $\phi_S$ defined here are {\em
opposite} to those defined in
Refs.~\cite{Mulders:1996dh,Boer:1998nt,Boer:1999uu}, which must be
taken into account when using expressions for azimuthal asymmetries from
these papers.\footnote{There is an inconsistency in Fig.~1 of Ref.~\cite{Boer:1999uu}
  and Fig.~1 of Ref.~\cite{Boer:1998nt}: 
according to the formulae given in those papers, 
the azimuthal angle shown in those figures (which is positive according to graphical
    convention \ref{conv1}) is equal to $-\phi$ and not to $\phi$.}

\section{Asymmetries and azimuthal moments}

Longitudinal single-spin asymmetries in lepton-proton scattering
should always be defined so that
\begin{equation}
A(\phi_h) \equiv \frac{d \sigma^{\to}(\phi_h) -
  d \sigma^{\gets}(\phi_h)}{d \sigma^{\to}(\phi_h) + 
  d \sigma^{\gets}(\phi_h)} ,
\end{equation}
where in the case of a beam spin asymmetry $d \sigma^{\to}$ refers to
positive helicity of the lepton.  In the case of a target spin
asymmetry $d \sigma^{\to}$ denotes target polarization {\em opposite}
to the direction either of the lepton beam or of the virtual
photon.\footnote{Note that target polarization opposite to the virtual
photon momentum corresponds to {\em positive} helicity of the proton
in the $\gamma^*p$ center of mass.}
{\em Azimuthal moments} associated with beam or target spin
asymmetries are defined as, e.g.\
\begin{equation}
\mbox{\large$\bm{\langle}$}\ensuremath{\sin \phi_h}
          \mbox{\large$\bm{\rangle}$}
 \equiv \frac{\int d \phi_h\; \sin \phi_h\; 
  \left[d \sigma^{\to}(\phi_h) - d \sigma^{\gets}(\phi_h)\right]}{
    \int d \phi_h\; \left[d \sigma^{\to}(\phi_h) +
                   d \sigma^{\gets}(\phi_h)\right] }
\end{equation}
and similarly for 
$\mbox{\large$\bm{\langle}$} \sin 2\phi_h \mbox{\large$\bm{\rangle}$}$ etc.  
As an alternative  
notation one may use $A^{\sin \phi_h} = 2 \mbox{\large$\bm{\langle}$} \sin
\phi_h \mbox{\large$\bm{\rangle}$}$.\footnote{In the literature sometimes the factor 2 is not
  included, a choice that we do not recommend.}  If the cross section is of the form
\begin{equation} \begin{split} 
\frac{d \sigma^{\to}}{d \phi_h} & = a_0 + a_1 \sin \phi_h, \\
\frac{d \sigma^{\gets}}{d \phi_h} & = a_0 - a_1 \sin \phi_h,
\end{split} \end{equation}  
then $A^{\sin \phi_h} = a_1/a_0$ has values between $-1$ and $+1$, as is
natural for an asymmetry.

The single spin asymmetry for transverse target polarization can be
written as
\begin{equation}
A(\phi_h, \phi_S) \equiv \frac{d \sigma(\phi_h, \phi_S) -
  d \sigma(\phi_h,\phi_S +\pi)}{d \sigma(\phi_h, \phi_S) + 
  d \sigma(\phi_h,\phi_S +\pi)} 
\end{equation}
and associated azimuthal moments as, e.g.\
\begin{equation}
\mbox{\large$\bm{\langle}$}\ensuremath{\sin(\phi_h+\phi_S)}
          \mbox{\large$\bm{\rangle}$}
 \equiv 
\frac{\int d \phi_h \; d\phi_S \;\sin
(\phi_h+ \phi_S) \; \left[d \sigma(\phi_h, \phi_S) -
  d \sigma(\phi_h,\phi_S +\pi)\right] }{\int d \phi_h\; d\phi_S\; 
\left[d \sigma(\phi_h, \phi_S) + d \sigma(\phi_h,\phi_S +\pi)\right] } 
\end{equation}
and similarly for 
$\mbox{\large$\bm{\langle}$} \sin(\phi_h-\phi_S) \mbox{\large$\bm{\rangle}$}$ 
etc.  It
should be straightforward to generalize these conventions to the case
of double spin asymmetries and of $|P_{h \perp}|$-weighted asymmetries~\cite{Boer:1998nt}.

\section*{Acknowledgments}

We thank the organizers and all the
participants of the workshop ``Transversity:~New Developments in Nucleon Spin
Structure'' (ECT$^\ast$, Trento, Italy, 14--18 June 2004). Special thanks
are due to M.~Anselmino, D.~Boer, A.~Metz, P.J.~Mulders, F.~Murgia, F.~Pijlman,
M.~Radici, and G.~Schnell 
for valuable input to the manuscript. 
This work is part of the EU Integrated Infrastructure Initiative
``Study of strongly interacting matter (HadronPhysics)'' under contract number
RII3-CT-2004-506078.

\bibliographystyle{apsrev}
\bibliography{mybiblio}

\end{document}